# UNITARIZABLE HIGHEST WEIGHT REPRESENTATIONS FOR AFFINE KAC-MOODY ALGEBRAS

Juan J. García-Escudero and Miguel Lorente.
Departamento de Física, Universidad de Oviedo
33007 Oviedo, Spain.

## Abstract

The main purpose of this work is to review the results obtained recently concerning the unitarization of highest weight representations for affine Kac-Moody algebras following the work of Jakobsen and Kac.

## 1. Introduction

In the past two decades interest on infinite dimensional Lie algebras has increased for both mathematicians and physicists.

The main purpuse of this work is to review the results obtained recently concerning the unitarization of highest weight representations for affine Kac-Moody algebras following the works of Jakobsen and Kac.

## 2. The Affine Kac-Moody Algebras

Let $\dot{g}$ be a finite dimensional semisimple complex Lie algebra with Chevalley basis $\{H_\alpha, E_\alpha, F_\alpha\}$ with $\alpha$ belonging to the set of simple roots. The elements of the so-called Cartan matrix $A$ are defined by

$$A_{jk} = \alpha_k(H_{\alpha j}) = \frac{2(\alpha_k, \alpha_j)}{\alpha_j, \alpha_j} \quad, \quad j, k = 1, 2, \ldots l$$

where $(\alpha_j, \alpha_k) = B(h_{\alpha_j}, h_{\alpha_k})$, $B(\,,\,)$ being the Killing form of $\dot{g}$ and $h_\alpha$ an element of the Cartan subalgebra $h$ which is assigned uniquely to each root $\alpha \in \Delta$ by the requirement that $B(h_\alpha, h) = \alpha(h)$ for all $h \in h$

The elements in the Chevalley basis satisfy

$$\left[H_{\alpha_j}, E_{\alpha_k}\right] = A_{jk} E_{\alpha_k} \quad, \quad \left[H_{\alpha_j}, F_{\alpha_k}\right] = -A_{jk} F_{\alpha_k}$$

$$\left[E_{\alpha_j}, F_{\alpha_k}\right] = \delta_{jk} H_{\alpha_j} \quad \text{for all } \alpha_j, \alpha_k \in \Delta$$

Every finite dimensional semisimple complex Lie algebra can be constructed from its Cartan matrix $A$ which satisfies the following propierties.

a)  $A_{ii} = 2 \quad \forall i \quad, \quad i = 1, \ldots, l,$

b)  $A_{ij} = 0, -1, -2, \text{or} -3 \quad \text{if} \quad i \neq j \quad, \quad i, j = 1, \ldots, l,$

c)  $A_{ij} = 0 \quad \text{if and only if} \quad A_{ji} = 0$

d)  det $A$ and all proper principal minors of $A$ are positive.

The starting point on the construction of infinite dimensional Kac-Moody algebras is the definition of a generalized Cartan-Matrix with elements $A_{ij}\ (i,j \in I\ ,\ I = 0, 1 \ldots, l)$ satisfying

a)  $A_{ii} = 2 \quad \forall i \in I$

b)  for $i \neq j$ $A_{ij}$ is either zero or a negative integer

c)  $A_{ij} = 0 \quad \text{if and only if} \quad A_{ji} = 0$

A Lie algebra whose Cartan matrix is a generalized Cartan matrix is called a Kac-Moody algebra[1],[2]. A Kac-Moody algebra is affine if its generalized Cartan matrix is such that det $A = 0$ and all the proper principal minors of $A$ are positive. In the following we restrict ourselves to the affine case.

For a Kac-Moody algebra the Cartan subalgebra $h$ is divided into two parts $h = h' \oplus h''$.

The basis elements of h' are $H_{\alpha_j}$ $(j \in I)$ and h" is the one-dimensional complementary subspace of h' in h generated by the element $d$.

The center $C$ in the affine case is one-dimensional. Every element of $C$ is a multiple of $h_\delta$ where $\delta$ is defined by the following condition

$$\delta(h) = 0 \quad \text{for } h \in h'$$
$$\delta(d) = 1 \quad \text{for } d \in h''$$

The Cartan subalgebra $h$ has dimension $l + 2$. In order to obtain a basis for $h^*$ we need $l+2$ linear functionals. We have $\alpha_k$ ($k = 0, \ldots l$), then we must define another linear functional $\Lambda_0$:

$$(\Lambda_0, \alpha_k) = \begin{cases} \frac{1}{2}(\alpha_0, \alpha_0) & \text{if } k = 0 \\ 0 & \text{if } k = 1, 2, \ldots l \end{cases}$$

$$(\Lambda_0, \delta) = \frac{1}{2}(\alpha_0, \alpha_0)$$

There exists two types of affine complex Kac-Moody algebras: Untwisted and Twisted. The untwisted ones $g^{(1)}$ may be constructed starting from any simple complex Lie algebra. The twisted affine Kac-Moody algebras $g^{(q)}$ ($q = 2,3$) can all be constructed as subalgebras of certain of these untwisted algebras.

### i) Untwisted affine Kac-Moody algebras

Let $\dot{g}$ be a simple complex Lie algebra of rank $l$. A realization of the complex untwisted affine Kac-Moody algebra $g^{(1)}$ is given by

$$g^{(1)} = \mathbb{C}c \oplus \mathbb{C}d \oplus \sum_{j \in \mathbb{Z}} z^j \otimes \dot{g}$$

with the following conmutation relations

$$\left[z^j \otimes a, z^k \otimes b\right] = z^{j+k} \otimes [a, b] + j\delta_{j,-k} B(a, b)c \quad \forall a, b \in \dot{g}$$
$$\left[z^j \otimes a, c\right] = 0$$
$$\left[d, z^j \otimes a\right] = j \, z^j \otimes a$$
$$[d, c] = 0$$

The $l+2$ –dimensional Cartan subalgebra is

$$\eta^{(1)} = (\mathbb{C}c) \oplus (\mathbb{C}d) \oplus \sum_{k=1}^{l} \mathbb{C}\left(1 \otimes h_{\alpha k}\right)$$

We can decompose $\eta^{(1)} = \eta' \oplus \eta''$ with $\eta' = (\cent c) \oplus \sum_{k=1}^{l} \cent\left(1 \otimes h_{\alpha_k}\right)$ and $\eta'' = \cent d$. A basis for $\eta'^*$ can be constructed with the elements $\alpha_k$ (corresponding to $H_{\alpha_k} \equiv 1 \otimes h_{\alpha_k}$  $j = 1, \ldots l$) and $\delta = \alpha_0 - \gamma_r$ (corresponding to the center generator $c$) where $\gamma_r$ is the extension of the highest root. We can take as basis element for $\eta''^*$ the linear functional $\Lambda_0$ (corresponding to the element $d$).

In this way the Killing form $B^{(1)}(\,,\,)$ for the untwisted case is

$$B^{(1)}\left(H_{\alpha_j}, H_{\alpha_k}\right) = \left(\alpha_j, \alpha_k\right) \quad j, k = 1, \ldots, l$$

$$B^{(1)}(c, H_{\alpha_k}) = (\delta, \alpha_k) = 0 \quad k = 1, \ldots, l$$

$$B^{(1)}(c, c) = (\delta, \delta) = 0$$

$$B^{(1)}(d, d) = (\Lambda_0, \Lambda_0) = 0$$

$$B^{(1)}(d, H_{\alpha_j}) = (\Lambda_0, \alpha_j) = 0 \quad j = 1, \ldots, l$$

$$B^{(1)}(d, c) = \frac{2(\Lambda_0, \alpha_0)}{(\alpha_0, \alpha_0)} = \frac{2(\Lambda_0, \delta)}{(\alpha_0, \alpha_0)} = 1$$

The set of roots can be divided into two subsets: real roots (satisfying $(\alpha,\alpha) > 0$) and imaginary roots (with $(\alpha,\alpha) \leq 0$)

**a) Real Roots:**

i) Extensions of positive roots on $\dot{g}$ i.e.

$$\alpha\left(1 \otimes h_{\alpha_k}\right) = \alpha\left(h_{\alpha_k}\right) \quad k = 1, 2, \ldots, l$$

$$\alpha(c) = \alpha(d) = 0$$

ii) Roots $j\delta + \alpha$, $j = \pm 1, \pm 2, \ldots$ is the extension of any non-zero root $\alpha$ of $\dot{g}$.

The basis elements of $g_{j\delta+\alpha}$ can be taken as $e_{j\delta+\alpha} = z^j \otimes e_\alpha$  $j = 0, \pm 1, \pm 2, \ldots$ and where $e_\alpha$ is an element of the Lie algebra $\dot{g}$ in a Cartan-Weyl basis. We have $\dim g_{j\delta+\alpha} = 1$

**b) Imaginary Roots:** $j\delta$, $j = \pm 1, \pm 2, \ldots$

The basis elements of $g_{j\delta}$ are $e^k_{j\delta} = z^j \otimes h_{\alpha_k}$  $k = 1, \ldots l$, $j = \pm 1, \pm 2, \ldots$. In this case $\dim g_{j\delta} = l$.

The complete set of Untwisted affine Kac-Moody algebras is (see Appendix).

$$A_l^{(1)} \ (l = 1, 2, \ldots) \ , \ B_l^{(1)} \ (l = 3, 4, \ldots) \ , \ C_l^{(1)} \ (l = 2, 3, \ldots)$$
$$D_l^{(1)} \ (l = 4, 5, \ldots) \ , \ E_6^{(1)} \ , \ E_7^{(1)} \ , \ E_8^{(1)} \ , \ F_4^{(1)} \ , \ G_2^{(1)}$$

### ii) Twisted Affine Kac-Moody algebras

Let $\dot{g}$ be a simple complex Lie algebra and $\tau$ a rotation of the set of roots of $\dot{g}$. If the rotation $\tau$ is not an element of the Weyl group of $\dot{g}$ then there exist an associated outer automorphism $\psi_\tau$ such that $\psi_\tau(h_\alpha) = h_{\tau(\alpha)}$. We have $\tau^q = 1$ and also $(\psi_\tau)^q = 1$ with $q = 2,3$. The eigenvalues of $\psi_\tau$ are $e^{2\pi i p/q}$, $p = 0, 1, \ldots q - 1$. Let $\dot{g}_p^{(q)}$ be the eigenspace corresponding to the eigenvalue $e^{2\pi i p/q}$. The Twisted Affine Kac-Moody algebra is then:

The sets of real and imaginary roots are in this case:

a) Real Roots: $j\delta + \alpha$, $j \mod q = p$ and $\alpha$ extensions of elements of $\dot{\Delta}_p^{(q)}$, $p = 0, 1, \ldots, q - 1$

b) Imaginary Roots: $j\delta$, $j = \pm 1, \pm 2, \ldots$

The Twisted Kac-Moody algebras can be labeled in the following way (see Appendix):

$$A_{2l}^{(2)} \ (l \geq 1) \quad A_{2l-1}^{(2)} \ (l \geq 3) \quad D_{l+1}^{(2)} \ (l \geq 2)$$
$$E_6^{(2)} \quad D_4^{(3)}$$

## 3. Highest Weight Representations. The Contravariant Hermitian Form

We can take a basis for an untwisted affine Kac-Moody algebra $g^{(1)}$ as ([1]):

$$e_0 = z \otimes F_{\gamma_r} \quad f_0 = z^{-1} \otimes E_{\gamma_r} \quad h_0 = \frac{2}{(\gamma_r, \gamma_r)} c - 1 \otimes H_{\gamma_r}$$
$$e_i = 1 \otimes E_i \quad f_i = 1 \otimes F_i \quad h_i = 1 \otimes H_i \quad i = 1, \ldots, l \ ,$$

where $\gamma_r$ is the highest positive root. From this basis we can construct

A subset $\Delta_+$ of $\Delta$ is called a set of positive roots if the following propierties are satisfied

$$g^{(1)} = \mathcal{C}c \oplus \mathcal{C}d \oplus \sum_{j \in Z} \left(z^j \otimes \dot{g}\right)$$

i) If $\alpha, \beta \in \Delta_+$ and $\alpha + \beta \in \Delta$ then $\alpha + \beta \in \Delta_+$

ii) If $\alpha \in \Delta$ then either $\alpha$ or $-\alpha$ belongs to $\Delta_+$

iii) If $\alpha \in \Delta_+$ then $-\alpha \notin \Delta_+$

For each set of positive roots we may construct a Borel subalgebra $b = \bigoplus_{\alpha \in \Delta_+ \cup \{0\}} g_\alpha$. A subalgebra $p \subset g$ such that $b \subset p$ is called a parabolic subalgebra.

Let $U(g)$ denote the universal enveloping algebra of $g$ and let $\omega$ be an antilinear anti-involution of $g$ $\left(i.e\ \omega[x, y] = [\omega y, \omega x]\ \text{and}\ \omega(\lambda x) = \overline{\lambda}\,\omega(x)\right)$ such that
$$g = p + \omega p$$

An antilinear anti-involution $\omega$ of $g$ is called consistent if $\forall\ \alpha \in \Delta$, $\omega g_\alpha = g_{-\alpha}$. When $\omega e_i = F_i$ and $\omega h_i = h_i$ $(i = 0, \ldots l)$ then $\omega$ is called the compact antilinear anti-involution and is denoted by $\omega_c$.

Let now $\Lambda : p \to \mathbb{C}$ be a 1-dimensional representation of $p$. A representation $\Pi : g \to gl(V)$ is called a highest weight representation with highest weight $\Lambda$ if there exists a vector $\vartheta_\Lambda \in V$ satisfying

  a) $\Pi(u(g))\vartheta_\Lambda = V$

  b) $\Pi(x)\vartheta_\Lambda = \Lambda(x)\vartheta_\Lambda \quad \forall x \in p$

An Hermitian form $H$ on $V$ such that
$$H(\vartheta_\Lambda, \vartheta_\Lambda) = 1$$
$$H(\Pi(g)u, v) = H(u, \Pi(\omega g)v) \quad \forall g \in g\ ;\ u, v \in V$$
is called contravariant. When $H$ is positive semi-definite, $\Pi$ is said to be unitarizable.

In the following we construct the Hermitian form $H$ ([4]). We choose a subspace $n \subset g$ such that $g = p \oplus n$. Then we have $U(g) = n\,U(g) \oplus U(p)$. Let $\beta$ be the proyection on the second sumand. Let $\Lambda$ be a 1-dimensional representation of a parabolic subalgebra $p$ (in particular a Borel subalgebra $b$ as in the integrable representations case) satisfying $\Lambda(\beta(u)) = \overline{\Lambda(\beta(\omega u))}\ \forall u \in U(g)$.

Let $p^\Lambda = \{x \in p\ /\ \Lambda(x) = 0\}$. The space
$$M_{p,\omega}(\Lambda) = U(g)/U(g)p^\Lambda$$

defines a representation of $g$ on $M_{p,\omega}(\Lambda)$ via left multiplication that is called a (generalized) Verma module and that is a highest weight representation. In addition it can be shown that there exists a unique contravariant hermitian form defined by

$$H(u, v) = \Lambda(\beta(\omega(v))u) \quad \text{for} \quad u, v \in U(g)$$

which is independent of the choice of $p$.

Let $I(\Lambda)$ denote the Kernel of H on $M_{p,\omega}(\Lambda)$. Then $H$ is nondegenerate on the highest weight module

$$L_{p,\omega}(\Lambda) = M(\Lambda)/I(\Lambda)$$

In the following we will give for each of the unitarizable representations (integrable, elementary and exceptional) the choice of $p$ and $\omega$ for which the hermitian form is nondegenerate and positive definite.

## 4. Integrable representations

Let $\Pi^{st} = \{\alpha_0, \ldots \alpha_l\}$ be the standard set of simple roots (one possible realization is given in the Appendix). The standard set of positive roots is $\Delta_+^{st} = \{\sum k_i \alpha_i / k_i = 0, 1, 2, \ldots \alpha_i \in \Pi^{st}\}$ and the corresponding Borel subalgebra is denoted by $b^{st}$:

$$b^{st} = \mathcal{c} c \oplus (1 \otimes \dot{b}) \oplus (z \otimes \dot{g}) \oplus (z^2 \otimes \dot{g}) \oplus \ldots =$$
$$+ \text{span}\{z^k \otimes h_i / k \geq 0, i = 0, \ldots l\} \oplus \text{span}\{z^k \otimes e_i / k \geq 0, i = 0, \ldots l\} \oplus$$
$$\oplus \text{span}\{z^k \otimes f_i / k > 0, i = 0, \ldots l\}$$

Let $\omega = \omega_c$ and let $\Lambda : b^{st} \to \mathcal{c}$ be a 1-dimensional representation of $b^{st}$ defined by

$$\Lambda(e_i) = 0 \quad \Lambda(h_i) = m_i \in Z_+ \quad (i = 0, \ldots l)$$

These representations are called the integrable highest weight representations. In particular if $g$ is finite-dimensional, these are the finite dimensional representations. The fundamental weight $\Lambda_0, \Lambda_1, \ldots \Lambda_l$ are such that

$$\Lambda_j(H_k) = \frac{2(\Lambda_j, \alpha_k)}{(\alpha_k, \alpha_k)} = \delta_{jk}$$

$$\Lambda_j(d) = 0 \qquad j, k = 0, \ldots l$$

In this way given the fundamental weight of a finite dimensional Lie algebra $\dot{g}$

$$\dot{\Lambda}_j(H_k) = \frac{2(\dot{\Lambda}_j, \alpha_k)}{(\alpha_k, \alpha_k)} = \delta_{jk} \quad j, k = 1, \ldots l$$

we can construct the fundamental weights of the Kac-Moody algebra $g$ as extensions of the fundamental weights $\dot{\Lambda}$ in the following way:

$$\Lambda_j = \dot{\Lambda}_j + \mu_j \Lambda_0$$

$$\mu_j = -\sum_{k=1}^{l} A_{ok} \left((\dot{A})^{-1}\right)_{kj}$$

$\dot{A}$ being the Cartan matrix of $\dot{g}$ and $\Lambda_0$ the linear functional defined in paragraph 2.

Every integrable highest weight representation is unitarizable.

## 5. Elementary representations

We know that for a finite dimensional simple Lie algebra $\dot{g}$ an infinite dimensional highest weight representation is unitarizable only if $\dot{\omega}$ is a consistent antilinear anti-involution corresponding to a hermitian symmetric space (see [6] – [9]).

The remaining unitarizable representations can be constructed only for Kac-Moody algebras related to these type of finite dimensional Lie algebras.

Let $\dot{b} = \dot{\eta} \oplus \bigoplus_{\alpha \in \dot{\Delta}_+ \cup \{0\}} \dot{g}_\alpha$ be a Borel subalgebra of the finite dimensional Lie algebra $\dot{g}$.

Consider the parabolic subalgebra (called "natural")

$$p^{nat} = z^n \otimes \dot{b} = \text{span}\left\{z^n \otimes h_i, z^n \otimes e_i\right\} \quad n \in Z, \quad i = \ldots 1$$

Take a Cartan decomposition of the Lie algebra $\dot{g}$ corresponding to a hermitian symmetric space:

$$\dot{g} = \dot{p}^- \oplus \dot{k} \oplus \dot{p}^+ \quad , \quad \dot{k} = \dot{k}^- \oplus \dot{\eta} \oplus \dot{k}^+$$

where $\dot{p}$ and $\dot{k}$ are the subspace and the subalgebra corresponding to non-compact and compact roots respectively.

We define an antilinear anti-involution $\omega$ of $g$ by

$$\omega(z^n \otimes k_i^+) = z^{-n} \otimes k_i^- \quad i = 2, \dots l \quad, \quad n \in Z$$

$$\omega(z^n \otimes p_1^+) = -z^{-n} \otimes p_1^- \qquad\qquad n \in Z$$

$$\omega(z^n \otimes h_i) = z^{-n} \otimes h_i \quad i = 0, 1, \dots l$$

where $k_2^+, \dots, k_l^+$ belong to $\dot{k}^+$ and where $p_1^+$ belongs to the root space corresponding to the unique simple non compact root. In the previous notation $e_1 = p_1^+$ and $e_i = k_i^+$ for i = 2, … l.

Consider now a set of highest weights $\Lambda_1, \dots, \Lambda_N$ corresponding to unitarizable highest weight modules for the hermitian symmetric space $\dot{g}$ (for an explicit calculation see [8] and [9]). Define a representation $p^{nat} \to ¢$ by

$$\Lambda(z^k \otimes x) = \sum_{i=1}^{N} C_i^k \Lambda_i(x)$$

for $x \in \dot{b}$ and $C_k^i \in ¢$ with $|C_k^i| = 1$

Then the resulting representation is called "elementary" and it is unitarizable.

## 6. Exceptional representations

Another class of unitary representations (called "exceptional") are constructed in this paragraph for the Kac-Moody algebra $z^k \otimes su(n,1)$ $k \in Z$, $n \geq 1$.

Let $\dot{g} = su(n,1)$ and let be a Cartan decomposition $\dot{g} = \dot{k} \oplus \dot{p}$. Then $\dot{\eta} = \dot{k}_1 \cap \dot{\eta} \oplus R\, \dot{h}_c$ where $\dot{k}_1 = [\dot{k}, \dot{k}]$ and $h_c$ belongs to the center of $\dot{g}$.

We take a realization of $g = z^k \otimes su(n,1)$ in terms of matrices $(a_{ij}(z))$ $i, j = 0, \dots n$. The matrix elements are of the form $a(z) = \sum_{n \in Z} a_n\, e^{in\theta}$ with $z = e^{i\theta}$. We will use the notation $\bar{a}(z) = \sum_{n \in Z} \bar{a}_n\, e^{in\theta}$. Let $p = \{(a_{ij}(z)) \in g\, / a_{ij} = 0 \quad \text{if}\ i > j\}$ be the parabolic subalgebra. The antilinear anti-involution acts in this case as

$$\omega(z^k \otimes h_c): \quad \omega(a_{00}(z)) = \bar{a}_{00}(z^{-1})$$

$$\omega(z^k \otimes \dot{p}^+): \quad \omega(a_{0j}(z)) = -\bar{a}_{j0}(z^{-1}) \quad j = 1, \dots n$$

$$\omega(z^k \otimes \dot{k}_1): \quad \omega(a_{ij}(z)) = \bar{a}_{ji}(z^{-1}) \quad i, j = 1, \dots n$$

Define a representation $\Lambda: p \to ¢$ by

$$\Lambda\left(a_{ij}(z)\right) = 0 \qquad i,j = 1, \ldots n$$

$$\Lambda\left(a_{0j}(z)\right) = 0 \qquad j = 1, \ldots n$$

$$\Lambda(a_{00}(z)) = -\int_{s^1} a_{00}(e^{i\theta})\, d\mu(\theta) = -\varphi(a_{00}(z))$$

where $\mu(\theta)$ is a positive mesure defined in the unit circle $s^1$ and infinitely supported.

It can be shown (see [4]) that the hermitian form $H$ (we remind that it is completely determined by giving $\omega$, $p$ and a representation of $p$) is positive definite and then the corresponding representation in the space $L_{p,\omega}(L)$ is unitary.

## 7.

The only remaining possibility in order to complete the set of all unitarizable representations for affine Kac-Mooey algebras is the corresponding to the highest component of a tensor product of an elementary with an exceptional representation for $z^k \otimes su(n,1)$ (see [5]).

# APPENDIX

In the following we give the set of simple roots for the affine Kac-Moody algebras in a Cartesian coordinate basis having in mind that the scalar product induced by the Killing form in the space $\eta^*$ is not euclidean ([3]):

Given $v = v^i e_i$  $w = w^i e_i$  $i = 1, \ldots n$.

$$(v, w) = \sum_{i=1}^{n-2} v^i w^i + v^{n-1} w^n + v^n w^{n-1}$$

where $e_i$ denoter the basis vector with 1 in the position $i$ and zero elsewhere.

## Untwisted Affine Kac-Moody Algebras

$A_1^{(1)}$:

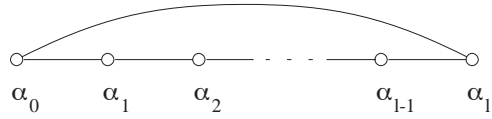

$\alpha_0 = -e_1 + e_2 + e_4$

$\alpha_1 = e_1 - e_2$

$\delta = e_4$

$A_l^{(1)}$ $(l \geq 2)$:

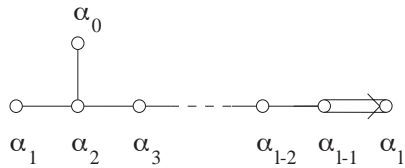

$\alpha_0 = -e_1 + e_{l+1} + e_{l+3}$

$\alpha_i = e_i - e_{i+1}$  $i = 1, \ldots l$

$\delta = e_{l+3}$

$B_l^{(1)}$ $(l \geq 3)$:

$\alpha_0 = -e_1 - e_2 + e_{l+2}$

$\alpha_i = e_i - e_{i+1}$  $i = 1, \ldots l-1$

$\alpha_l = e_l$

$\delta = e_{l+2}$

$C_l^{(1)}$ $(l \geq 2)$:

$$\alpha_0 = -2e_1 + e_{l+2}$$
$$\alpha_i = e_i - e_{i+1} \quad i = 1, \ldots l-1$$
$$\alpha_l = 2e_l$$
$$\delta = e_{l+2}$$

$D_l^{(1)}$ $(l \geq 4)$:

$$\alpha_0 = -e_1 - e_2 + e_{l+2}$$
$$\alpha_i = e_i - e_{i+1} \quad i = 1, \ldots l-1$$
$$\alpha_l = e_{l-1} + e_l$$
$$\delta = e_{l+2}$$

$E_6^{(1)}$

$$\alpha_0 = \frac{1}{2}(-e_1 - e_2 - e_3 - e_4 - e_5 + e_6 + e_7 - e_8 + 2e_{10})$$
$$\alpha_1 = \frac{1}{2}(e_1 - e_2 - e_3 - e_4 - e_5 - e_6 - e_7 + e_8)$$
$$\alpha_i = -e_{i-1} + e_i \quad i = 2, 3, 4, 5$$
$$\alpha_6 = e_1 + e_2$$
$$\delta = e_{10}$$

$E_7^{(1)}$

$$\alpha_0 = e_7 - e_8 + e_{10}$$

$$\alpha_1 = \frac{1}{2}(e_1 - e_2 - e_3 - e_4 - e_5 - e_6 - e_7 + e_8)$$
$$\alpha_i = -e_{i-1} + e_i \quad i = 2, 3, 4, 5, 6$$
$$\alpha_7 = e_1 + e_2$$
$$\delta = e_{10}$$

$E_8^{(1)}$ 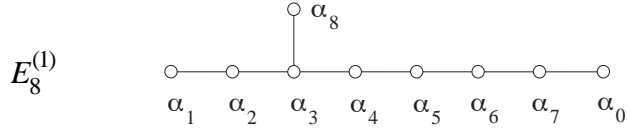

$$\alpha_0 = -e_7 - e_8 + e_{10}$$
$$\alpha_1 = \frac{1}{2}(e_1 - e_2 - e_3 - e_4 - e_5 - e_6 - e_7 + e_8)$$
$$\alpha_i = -e_{i-1} + e_i \quad i = 2, 3, 4, 5, 6, 7$$
$$\alpha_8 = e_1 + e_2$$
$$\delta = e_{10}$$

$F_4^{(1)}$ 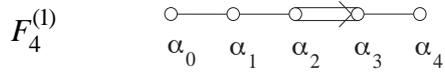

$$\alpha_0 = -e_1 - e_2 + e_6$$
$$\alpha_i = e_{i+1} - e_{i+2} \quad i = 1, 2$$
$$\alpha_3 = e_4$$
$$\alpha_4 = \frac{1}{2}(e_1 - e_2 - e_3 - e_4)$$
$$\delta = e_6$$

$G_2^{(1)}$ 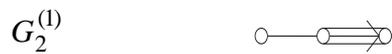

$$\alpha_0 = e_1 + e_2 - 2e_3 + e_5$$
$$\alpha_1 = -2e_1 + e_2 + e_3$$
$$\alpha_2 = e_1 - e_2$$
$$\delta = e_5$$

## Twisted Affine Kac-Moody Algebras

$A_2^{(2)}$ 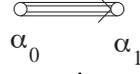

Outer automorphism

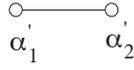

$\tau(\alpha'_1) = \alpha'_2$

$\tau(\alpha'_2) = \alpha'_1$

Roots

$\alpha_0 = -e_1 + e_3 + \frac{1}{2}e_5$

$\alpha_1 = \frac{1}{2}(e_1 - e_3)$

$\delta = \frac{1}{2}e_5$

$A_{2l}^{(2)} \ (l \geq 2)$ 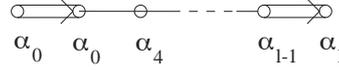

Outer automorphism

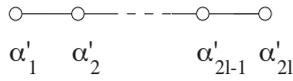

$\tau(\alpha'_k) = \alpha'_{2l+1-k}$

$k = 1, 2, \ldots l$

Roots

$\alpha_0 = -e_1 + e_{2l+1} + \frac{1}{2}e_{2l+3}$

$\alpha_i = \frac{1}{2}(e_i - e_{i+1} + e_{2l-i+1} - e_{2l-i+2}) \quad i = 1, \ldots, l-1$

$\alpha_l = \frac{1}{2}(e_l - e_{l+2})$

$\delta = \frac{1}{2} e_{2l+3}$

$A_{2l-1}^{(2)} \ (l \geq 3)$ 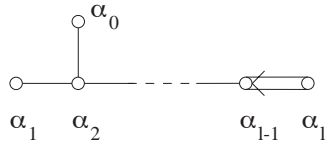

Outer automorphism

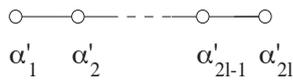

$\tau(\alpha'_k) = \alpha'_{2l-k} \quad k = 1, \ldots, l-1$

$\tau(\alpha'_l) = \alpha'_l$

Roots

$\alpha_0 = \frac{1}{2}(-e_1 - e_2 + e_{2l-1} + e_{2l} + e_{2l+2})$

$\alpha_i = \frac{1}{2}(e_i - e_{i+1} + e_{2l-i} - e_{2l-i+1}) \quad i = 1, \ldots, l$

$\alpha_l = e_l - e_{l+1}$

$\delta = \frac{1}{2} e_{2l+2}$

$D_{l+1}^{(2)}\ (l \geq 2)$ 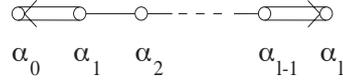

| Outer automorphism | Roots |
|---|---|
| 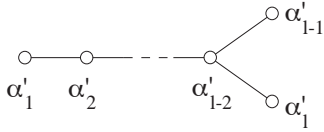 $\tau(\alpha'_k) = \alpha'_k \quad k = 1, 2, \ldots, l-1$ $\tau(\alpha'_l) = \alpha'_{l+1}$ $\tau(\alpha'_{l+1}) = \alpha'_l$ | $\alpha_0 = -e_1 + \frac{1}{2} e_{l+3}$ $\alpha_i = e_i - e_{i+1} \quad i = 1, \ldots, l-1$ $\alpha_l = e_l$ $\delta = \frac{1}{2} e_{l+3}$ |

$E_6^{(2)}$ 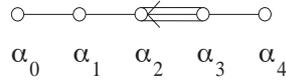

| Outer automorphism | Roots |
|---|---|
| 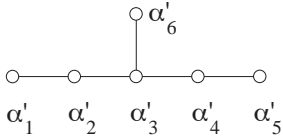 $\tau(\alpha'_k) = \alpha'_{6-k} \quad k = 1, 2, \ldots, 5$ $\tau(\alpha'_6) = \alpha'_6$ | $\alpha_0 = \frac{1}{2}(-e_5 + e_6 + e_7 - e_8 + e_{10})$ $\alpha_1 = \frac{1}{4}(e_1 - e_2 - e_3 - 3e_4 + e_5 - e_6 - e_7 + e_8)$ $\alpha_2 = \frac{1}{2}(-e_1 + e_2 - e_3 + e_4)$ $\alpha_3 = -e_2 + e_3$ $\alpha_4 = e_1 + e_2$ $\delta = \frac{1}{2} e_{10}$ |

$D_4^{(3)}$ 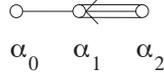

$\alpha_0 \quad \alpha_1 \quad \alpha_2$

| Outer automorphism | Roots |
|---|---|
| 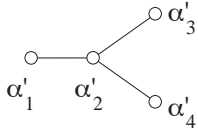 $\alpha'_1 \quad \alpha'_2 \quad \alpha'_3 \quad \alpha'_4$ $\tau(\alpha'_1) = \alpha'_3$ $\tau(\alpha'_2) = \alpha'_2$ $\tau(\alpha'_3) = \alpha'_4$ $\tau(\alpha'_4) = \alpha'_1$ | $\alpha_0 = \frac{1}{3}(-2e_1 - e_2 - e_3 + e_6)$ $\alpha_1 = \frac{1}{3}(e_1 - e_2 + 2e_3)$ $\alpha_2 = e_2 - e_3$ $\delta = \frac{1}{3} e_6$ |